# Solar system and small-field astrometry

Erik Høg (Niels Bohr Institute, Copenhagen, Denmark)
George Kaplan (US Naval Observatory, Washington DC, USA, retired)

2017.08.03 - 4 links to dropbox changed to my website.
2014.11.06 – with comments from J.-E. Arlot and P. Tanga, updates in red of the version v2 of 2014.09.28, including comments by M. Shao.

**ABSTRACT:** Astrometric issues for solar system studies are discussed. An overview gives references and cover all aspects of the solar system where astrometry is important: orbits of planets, moons, asteroids and NEOs, masses of asteroids, occultations of asteroids and KBOs, and families of asteroids and KBOs. The roles of astrometry from the ground, from Gaia and from a Gaia successor are discussed, but *not* small-field astrometry *from space*. It appears from work with CCD cameras at the 1.55 m astrometric reflector in Flagstaff that an accuracy of 1 mas is the best possible from the ground during one night observing when using ordinary telescopes, i.e. without wave-front correctors, and for field sizes larger than 2 arcmin. It has been seen that the same accuracies can be reached with the much larger 4-m class telescope on Hawaii although it is not specifically designed for astrometry. The accuracy of 1 mas from the ground refers mainly to non-moving point sources, but it is expected that 1 mas can be reached from the ground for solar system bodies from many nights of observations when phase effects are taken into account.

## Content

This paper is a contribution to the study of a Gaia successor by Høg (2014).
Section 1 is a slightly modified version of a section in Høg (2014).
Section 2 is an overview of the solar system issues by George Kaplan.

## 1. Solar system and small-field astrometry

Astrometric issues for solar system studies were discussed in July to September 2014 in correspondence with a number of colleagues. Paolo Tanga (2014) defines three levels of astrometric accuracy in order to show the increasing amount of science for the solar system obtained by better accuracy and he says that 0.1 mas would be useful for four specific scientific purposes. Tanga expects that 0.1 mas will only be obtained with observations from space and that 1 mas will be possible from the ground. Michael Shao (2014) comments on the report by Paolo Tanga and disagrees on some points which could, regrettably, not be resolved in the correspondence.

A new reduction of old astrometric observations of solar system objects will according to Arlot (2014) be obtained when the Gaia reference star catalogue will be available. Its accuracy will give an increase of the accuracy of the many old observations obtained since photography was introduced about 1890. A Gaia successor will secure high-precision astrometry in the solar system also in the far future and it appears that a measurement accuracy of 1 mas will be sufficient because the irregularity of the figure of the objects will set a limit, according to Arlot.

But Tanga (2014) claims that 0.1 mas would be useful and this corresponds e.g. to 150 m at a distance of 2 AU. So that kind of accuracy could be important for the smaller objects (most of the main-belt asteroids and NEOs) as Tanga points out, but for the larger objects, Arlot's concern about center-of-light vs. center-of-mass would come into play. Tanga discusses this effect and how it can be modeled and be taken into account. He concludes that the accuracy of 0.1 mas can probably be approached from space by the Gaia mission. Ground-based observations can probably reach the 1 mas accuracy for a mean position based on many nights of observations when the Gaia absolute reference frame becomes available. This requires that the shape of the object be determined by sufficient photometric observations. The Gaia frame will only remain sufficiently accurate during a few years after the Gaia mission as discussed below.

Overviews of issues for *future* solar system studies from two colleagues in the USA, George Kaplan (US Naval Observatory, Washington, retired) and Hugh Harris (US Naval Observatory, Flagstaff) are given here *in extenso* in section 2 and in Harris (2014). These overviews give references and cover, with some overlap, all aspects of the solar system where astrometry is important: orbits of planets, moons, asteroids and NEOs, masses of asteroids, occultations of asteroids and KBOs, and families of asteroids and KBOs. The roles of astrometry from the ground, from Gaia and from a Gaia successor are discussed by both.

Harris expects accuracies of 1-5 mas for ground-based observations with small-field astrometry when Gaia results become available in the form of a very accurate absolute reference frame with a large number of stars, close to one billion. The ground-based observations determine positions relative to the reference frame for other objects in the field, i.e. stars or solar system objects. For stars, the proper motions and parallaxes can be derived after years of observations. For solar system objects, orbits can be determined.

Consequently the same accuracies of 1-5 mas are expected for predictions of positions with the new orbits, representing improvement by a factor of 10-100 over the present. Harris and his colleagues now use a field size of typically 10 arcmin (i.e. a square with these sides), and reach standard errors of 3 mas for a single exposure, 2 mas if the reference field is 6 arcmin. He expects they will move toward smaller field sizes in the future.

A comparison with the expression by Lindegren (1980) for the astrometric errors due to the atmosphere is given at the end of Harris (2014). This expression indicates that the improvement with smaller fields goes with the field to the power 0.25, much more slowly than suggested by the numbers given by Harris. This issue should be further investigated, but on the basis of this finding and these reports I suggest that an accuracy of 1 mas is the best possible from the ground when using ordinary telescopes, i.e. without wave-front correctors.

If 1 mas is the limit, an accuracy (random error) of 0.5 mas is adequate for the reference stars since such a frame would contribute only 0.1-0.2 mas to the standard error of the object. It should be noted that the systematic errors with astrometry satellites are much smaller than the random errors, thus for Gaia systematic errors about 0.001 mas is expected. With ground-based astrometry systematic errors are often comparable to the random errors.

The reference frame should contain all stars to G=20 mag, but it need not go fainter for the sake of solar system work, as explained in Harris (2014). The limit of 1 mas is expected for observation of a star or solar system object in a reference frame *in a single night*. For stars, the accuracy can be improved by observations on many nights if the reference system is more accurate. But this is not required for a solar system object, since e.g. a KBO at a distance of 40 AU will move about 1.5 arcmin per day due to the Earth's motion and therefore soon appear among other reference stars.

The Gaia frame will have errors of 1.8 mas at G=20 mag in 2026, 3.5 mas in 2036, and 8.8 mas in 2066 as explained in Høg (2014). This could not at all satisfy solar system observers. It has been suggested to use the more accurate Gaia stars of G=16 and brighter, but they are too sparse to fill the small fields required to obtain the 1 mas accuracy in observations.

A solution to this problem would be a Gaia successor as advocated here. Another solution to the problem would be a densification of the optical reference frame as proposed in Høg (2014) by Zacharias. An all-sky survey with 1-meter class telescopes could give a frame with 2 mas accuracy of stars to G=22 mag which is however not the accuracy wanted for solar system work. Thus, a Gaia successor providing 0.5 mas accuracy (standard errors) or better at G=20 mag is required.

# 2. The solar system - an overview

collected by George Kaplan, dated 11-07-2014

Precise fundamental ephemerides of the planets in the solar system, computed by the Jet Propulsion Laboratory (JPL) in the U.S., l'Institut de Mécanique Céleste et de Calcul des Éphémérides (IMCCE) in France, and the Institute of Applied Astronomy (IAA) in Russia, rely mostly on radar and spacecraft data for the inner solar system, spacecraft data for Jupiter and Saturn, and lunar laser ranging for the Moon. The link to the ICRF is provided by VLBI observations of the spacecraft (Folkner and Border 2012). Precise optical astrometry is still needed for the planets beyond Saturn (Standish & Williams 2012) as well as for most asteroids and Trans-Neptunian Objects (TNOs). Infrequent spacecraft flybys of these objects are very useful for determining masses but do not constrain the orbital parameters very well. For the far outer planets, the modern astrometric technique is to observe the positions of the natural satellites and infer the position of the center of mass of the planet-satellite system. Better orbits for the satellites are also of interest for modeling the dynamical history of the satellite systems.

In the inner solar system, there is much current interest in near-Earth asteroids (NEOs) that pose some risk of colliding with Earth in the future. These are generally small, faint objects that do not have long-term (or any) observing histories. Estimating the future paths of these objects and their risk to Earth therefore requires accurate absolute astrometry, including that of very faint reference stars beyond the Gaia limit.

Asteroid masses can be determined by very precise astrometry before and after close encounters between asteroids (Hilton et al. 1996). However, asteroid-asteroid encounters that are useful in determining asteroid masses are infrequent events, typically years apart. Other techniques, including radar observations and better photometry, combined with better knowledge of asteroid taxonomic classes and families, are more likely to yield asteroid mass estimates in greater numbers. But determining the membership in and dynamical histories of the families requires continuing and improved astrometric observations. Unknown masses of asteroids contribute "noise" in the ephemerides of the planets in the inner solar system (Standish & Fienga 2002), although the last decade's worth of ranging data from Mars spacecraft have allowed for improvements in the determination of the masses of the asteroids that significantly perturb that planet's orbit (Kuchynka & Folkner 2013).

Murison & Harris (2014) have shown that the improved faint star positions from Gaia will considerably help to reduce the uncertainties in the paths of occultations of these stars by planets, asteroids, or TNOs, one of the techniques for measuring their size and shape.

Arlot (2014) has pointed out that the Gaia catalog will provide much improved reference star positions for old astrometric observations, even those on old photographic plates. He states that even for observations in the early 20[th] century, Gaia reference star positions should be accurate to several milliarcseconds, as faint as 16[th] magnitude, based on pre-launch estimates published by Mignard (2012). Re-reduction of these old observations could therefore provide much improved positions of many solar system objects going back decades. Such re-reductions, although labor intensive, have the potential to significantly improve the orbital parameters of many solar system bodies, to the limit imposed by uncertainties in the location of their centers of light (or centers of figure) with respect to their centers of mass. For asteroids, such re-reductions might also allow for refined mass estimates for those that have undergone encounters with other asteroids.

_______________________________________________

Submitted by G. Kaplan (USNO, retired) with contributions from W. Folkner (JPL) and J.-E. Arlot (IMCCE)